\pdfoutput=1
\documentclass[twocolumn,prd,superscriptaddress,preprintnumbers,nofootinbib]{revtex4}
\usepackage{graphicx}
\usepackage{epsfig}
\usepackage{lipsum}
\usepackage{mathrsfs}
\usepackage{enumitem}
\usepackage{bm}
\usepackage{latexsym,amssymb,amsmath,amsfonts,amssymb,txfonts,pxfonts,wasysym,float}
\usepackage{color}

\newcommand{\beq}[1]{\begin{equation}\label{#1}}
\newcommand{\eeq}{\end{equation}}
\newcommand{\bea}[1]{\begin{eqnarray} \label{#1}}
\newcommand{\eea}{\end{eqnarray}}
\newcommand{\ba}{\begin{array}}
\newcommand{\ea}{\end{array}}

\def\be{\begin{equation}}
\def\ee{\end{equation}}
\def\gs{\mathrel{
   \rlap{\raise 0.511ex \hbox{$>$}}{\lower 0.511ex \hbox{$\sim$}}}}
\def\ls{\mathrel{
   \rlap{\raise 0.511ex \hbox{$<$}}{\lower 0.511ex \hbox{$\sim$}}}}

\newcommand{\postscript}[2]{\setlength{\epsfxsize}{#2\hsize}
   \centerline{\epsfbox{#1}}}

\newcommand{\comment}[1]{}

\usepackage[usenames,dvipsnames]{xcolor}
\definecolor{orange}{cmyk}{0,0.5,1,0}
\definecolor{rossoCP3}{cmyk}{0,.88,.77,.40}
\definecolor{graa}{rgb}{0.8,0.8,0.8}
\definecolor{blaa}{rgb}{0.2,0.2,0.6}

\begin{document}

\title{\color{rossoCP3}{Oscillations of sterile neutrinos from dark matter decay eliminates the IceCube-Fermi tension}}

\author{Luis A. Anchordoqui}

\affiliation{Department of Physics and Astronomy,  Lehman College, City University of
  New York, NY 10468, USA
}

\affiliation{Department of Physics,
 Graduate Center, City University
  of New York,  NY 10016, USA
}

\affiliation{Department of Astrophysics,
 American Museum of Natural History, NY
 10024, USA
}

\author{Vernon Barger}

\affiliation{Department of Physics, University of Wisconsin, Madison, WI 53706, USA}

\author{Danny Marfatia}
\affiliation{Department of Physics and Astronomy, University of Hawaii, Honolulu, HI 96822, USA}

\author{Mary Hall Reno}

\affiliation{Department of Physics and Astronomy, University of Iowa, Iowa City, IA 52242, USA}

\author{Thomas J. Weiler}
\affiliation{Department of Physics and Astronomy, Vanderbilt University, Nashville TN 37235, USA}

\begin{abstract}
  \vskip 2mm \noindent IceCube has observed a flux of cosmic
  neutrinos, with a ``bump"  in the energy range $10 \lesssim
E/{\rm TeV} \lesssim 100$ that creates a
$3\sigma$ tension with $\gamma$-ray data from the Fermi
satellite. This has been interpreted as evidence for a population of
hidden cosmic-ray accelerators. We propose an alternative
explanation of this conundrum on the basis of cold dark matter which decays into sterile neutrinos that after oscillations produce the bump in the cosmic neutrino spectrum. 
\end{abstract}
\maketitle

\section{Introduction}

The most immediate message emerging from IceCube's discovery of cosmic
neutrinos is that the flux level observed is exceptionally high by
astronomical standards~\cite{Aartsen:2013jdh,Aartsen:2014gkd,Abbasi:2020jmh}. The magnitude of the observed diffuse neutrino flux is above about the level of the Waxman-Bahcall bound~\cite{Waxman:1998yy}, which applies to neutrino
production in optically thin sources. As always, the devil is in the details. Neutrinos are inevitably produced in association with
gamma rays 
when accelerated baryonic cosmic rays produce charged and neutral pions in
interactions with intense radiation fields or dense clouds of gas surrounding the
accelerator. The subsequent decay of charged pions via $\pi^+ \to
\mu^+ \nu_\mu$
 followed by muon decay  $\mu^+ \to e^+ \nu_e \overline \nu_\mu$
 (and the charge-conjugate processes) produce a neutrino flux,
 whereas the associated gamma-ray flux originates in the decay of neutral pions,
 $\pi^0 \to \gamma \gamma$. On average, pionic $\nu$'s and $\gamma$'s carry one quarter and one half of the energy of the parent pion,
 respectively.  If the sources were optically thin, then  on the basis of these
 approximations we would expect
 roughly equal fluxes of $\nu$'s and $\gamma$'s~\cite{AlvarezMuniz:2002tn}. Before confronting this
 equality with experiment, we must account for the fact that, unlike
 neutrinos, gamma rays are 
 degraded in energy by electromagnetic
 cascades and contribute to the diffuse GeV-TeV flux that has been precisely
 measured by the Fermi
satellite~\cite{Ackermann:2014usa}. The
neutrino flux observed by IceCube in the energy range $10 \lesssim E/{\rm TeV} \lesssim 100$ creates a
$3\sigma$ tension with  Fermi data where there isn't a commensurate gamma ray flux.

The IceCube-Fermi tension has been interpreted as evidence for a
population of hidden cosmic-ray
accelerators, {\it viz.} sources that are more efficient neutrino than
gamma-ray emitters~\cite{Murase:2015xka,Capanema:2020rjj,Capanema:2020oet}. This
interpretation, however, requires some fine-tuning as it would need
a source environment with:
\begin{itemize}[noitemsep,topsep=0pt]
\item~a low-density region for cosmic ray protons to be accelerated without
suffering catastrophic spallations;
\item~a medium-density region where the baryonic cosmic rays can
  interact, but charged pions are able to decay;
\item~a
high-density region in the outer parts of the source to trap the photons, and perhaps also
the baryonic cosmic rays.
\end{itemize}
Since the proton-proton cross section is
comparable to the pion-proton
cross section ($\sigma_{\pi p}/\sigma_{pp} \simeq 2/3$),  
the first two requirements are difficult to reconcile. Neutrino production requires an optically thin source: the high energy cutoff of the cosmic ray spectrum comes primarily from pion production,  however, pion
production cannot be significant at lower energies to allow
cosmic rays to be adequately accelerated. While protons experience magnetic confinement, neutrons can escape, and then decay to yield a cosmic ray proton  flux.

These considerations can be translated to conditions on the characteristic time scales:
the proton interaction time scale $\tau_{\rm int}$, the
neutron decay lifetime $\tau_{\rm n}$, the cycle time of
confinement $\tau_{\rm cycle}$, and the total proton confinement time
$\tau_{\rm conf}$.
For sufficient acceleration, $\tau_{\rm int} \gg \tau_{\rm cycle}$ is required. Additionally, in order for neutrons to escape the source, 
$\tau_{\rm n} > \tau_{\rm cycle}$. Finally, to produce neutrons and neutrinos, $\tau_{\rm int} \ll \tau_{\rm conf}$~\cite{Ahlers:2005sn}. These conditions are required of optically thin sources. A condition for hidden cosmic-ray accelerators is $\tau_{\rm n} < \tau_{\rm cycle}$,  which contradicts the conditions for neutrino emission from optically thin sources. Instead, to trap gamma rays (produced via $\pi^0$
decay) and cosmic rays, one would need a high-density optically
thick region in the outer parts of the source to encircle the
optically thin engine. This hypothetical source structure is certainly
not the most natural and it needs some fine-tuning.

In this paper we present an alternative
explanation of the IceCube-Fermi tension in which active neutrinos originate
via decaying dark matter into sterile neutrinos. These sterile neutrinos
oscillate to produce the ``bumpy" signal in the $10 \lesssim E/{\rm TeV} \lesssim 100$ energy range. The dark matter
origin of IceCube neutrinos has been extensively discussed in the
literature. However, in most of these studies the dark matter particle
couples to the Standard Model (SM) through: {\it (i)}~the quark $Q$ doublet, {\it (ii)}~the lepton $L$ and/or
Higgs $H$ doublets, {\it (iii)}~the $W$ or $Z$ gauge bosons, {\it
  (iv)}~a massless dark photon which mixes with the ordinary photon, or
{\it (v)}~a dark $Z'$ which mixes with the
$Z$~\cite{Feldstein:2013kka,Esmaili:2013gha,Bai:2013nga,Ema:2013nda,Bhattacharya:2014vwa,Ema:2014ufa,Rott:2014kfa,Anchordoqui:2015lqa,Bhattacharya:2017jaw,Chianese:2017nwe,Bhattacharya:2019ucd}. Therefore,
in these models, dark matter decay leads to fluxes 
of both neutrinos and photons. As a matter of fact, the photon signal has been used to
constrain these scenarios~\cite{Murase:2015gea,Ahlers:2015moa,Esmaili:2015xpa}.  The novelty of the scenario proposed herein is
that the dominant flux of IceCube neutrinos in the energy range $10
\lesssim E/{\rm TeV} \lesssim 100$ originates via oscillations of
sterile neutrinos produced at cosmological distances, but without a gamma-ray
counterpart, thereby addressing the
IceCube-Fermi tension; for earlier work that explains the IceCube PeV events without consideration of Fermi data see~\cite{Berezhiani:2015fba}. 

The layout of the paper is as follows.  We begin in Sec.~\ref{sec:2}
by reviewing the cosmological constraints on decaying dark matter. In Sec.~\ref{sec:3} we provide an outline of the basic
setup of the Boltzmann transport equation, specifying model assumptions
on the neutrino mass spectrum, sterile neutrino interactions, and
oscillation parameters. In Sec.~\ref{sec:4} we confront the model
predictions with IceCube data. In Sec.~\ref{sec:5} we discuss related phenomenology. Finally, we summarize our results and
draw our conclusions in Sec.~\ref{sec:6}.

\section{Constraints on decaying dark matter}
\label{sec:2}

A large number of observations in cosmology and astrophysics provide
overwhelming evidence for dark
matter~\cite{Bertone:2004pz}. However, so far the origin of this
evidence has been purely gravitational; hence we
have very few clues about the particle nature of the dark matter. 
We
know dark matter is essential for structure formation in the late
universe, so most of it (though not all) must be stable on cosmological time scales. 

The dominant paradigm in dark-matter phenomenology has been to
consider frameworks in which
the total dark matter (DM) density today, $\Omega_{\rm dm}$, is
made up of one stable particle species. However, it may well be that many
particle species -- perhaps even a vast
number~-- contribute nontrivially to the
abundance, with some of
these quasi-stable, as in Ref.~\cite{Dienes:2011ja}.

For simplicity, we consider two cold components to constitute dark matter, one of which is stable, denoted by $\chi_{\rm sc}$, and the other which decays, denoted by $\chi_{\rm dc}$, with a lifetime $\tau_{\rm {dc}} \agt t_{\rm
  LS}$, where  $t_{\rm LS}$ is the time of last scattering. 
  The fraction $F$ of decaying dark matter is
  \begin{equation}
  F \equiv\frac{ \Omega_{{\rm dc}}} { \Omega_{{\rm dm}}}\,,
  \end{equation}
  where $\Omega_{{\rm dm}} = \Omega_{{\rm dc}} + \Omega_{{\rm sc}}$, and the $\Omega$'s are the energy densities today in units of the critical density. Here, $\Omega_{{\rm dc}}$ is the density parameter today as if none of it had decayed.
  
We consider the particular case in which $\chi_{\rm dc}$ decays with a branching fraction of essentially
unity via $\chi_{\rm dc} \to \nu_s \bar \nu_s$, where $\nu_s$ is
a sterile neutrino that behaves as dark radiation. See Ref.~\cite{Berezhiani:2015fba}  for a model
that realizes this scenario. The dynamics associated with the energy of the $\nu_s$ could  change the evolution of cosmological perturbations,
leading to observable consequences,  thereby allowing us to place constraints
on $F$ and the decay width $\Gamma_{\rm dc}$ of $\chi_{\rm dc}$. Anisotropies in the cosmic microwave
background (CMB) temperature and polarization fields measured by the
Planck mission~\cite{Aghanim:2018eyx} when combined with Baryonic Acoustic Oscillations data from BOSS
data release 12 (aka DR-12)~\cite{Dawson:2012va} imply
\begin{equation}
\label{eq:flim}
  F < 1.28 \times 10^{-2}
\end{equation}
and
\begin{equation}
 F \ \Gamma_{\rm dc}  < 2.25 \times
  10^{-5}~{\rm Gyr}^{-1} 
\label{Gammaflimit}
\end{equation}
at the 95\% CL~\cite{Nygaard:2020sow}. This small fraction of dark radiation cannot alter significantly the
expansion
rate~\cite{Berezhiani:2015yta,Chudaykin:2016yfk,Poulin:2016nat,Chudaykin:2017ptd,Vattis:2019efj,Clark:2020miy,Abellan:2020pmw,Anchordoqui:2020djl}. Therefore, we can safely assume that the evolution of
the universe is described by the standard $\Lambda$ cold dark
matter cosmological model, with the Hubble parameter given by
\begin{equation}
H(z) = H_0 \sqrt{\Omega_\Lambda + \Omega_{m} (1+z)^3 + \Omega_{r}
  (1+z)^4} \, ,
\end{equation}
where $H_0 = 100 \ h~{\rm km/s/Mpc}$ is the Hubble constant and $\Omega_\Lambda$,
$\Omega_m$, and $\Omega_r$ are the present day fractions of the dark energy density, the
non-relativistic matter density, and the radiation density. The cosmological parameters are normalized to CMB data, with $h=0.6766$~\cite{Aghanim:2018eyx}.

\section{Boltzmann transport}
\label{sec:3}

The evolution of the neutrino density in phase space is driven by
Boltzmann's transport equation, which equates the directional derivative of the distribution function $f$ along the phase flow to a collision integral. Strictly speaking, the change in $f$ along a phase space trajectory with affine parameter $\lambda$ is equal to the phase space density $C[f]$ of point-like collisions that add or remove neutrinos from the trajectory:
\begin{equation}
\frac{df}{d\lambda} = C[f] \, .
\end{equation}
The phase space measure is defined in such a way that $f$ and $C[f]$ are both invariant scalars. For practical computations, it is necessary to introduce
phase space coordinates, which we take to be spacetime position coordinates $x^\mu$ and momentum space
coordinates $P^\mu$. Then,
\begin{equation}
\left(\frac{dx^\mu}{d \lambda} \frac{\partial }{\partial x^\mu} +
\frac{dP^\mu}{d\lambda} \frac{\partial }{\partial P^\mu} \right)
f(x^\mu,P^\mu)= C[f] \, .
\label{Boltzmann-coordinates}
\end{equation}
Before proceeding, we pause to present our notation. We adopt lower-case Greek
letters from the middle of the alphabet for spacetime components (where indices take on values $0, 1,
2,3$), lower-case Latin letters from the middle of the alphabet for spatial components (where
indices take on values $1, 2, 3$), lower-case Greek letters from the
beginning of the alphabet for neutrino flavors (where indices take on
value $e,\mu,\tau,s$), and lower-case Latin letters from the
beginning of the alphabet for neutrino mass eigenstates (where
indices take on values $1, 2, 3,4$).  Geometry comes into play via the geodesic equation specifying the neutrino trajectories: 
\begin{equation}
\frac{dP^\mu}{d\lambda} + \Gamma^\mu_{\nu \sigma} P^\nu P^\sigma = 0 \,,
\label{derivadacovariante}
\end{equation}
where
\begin{equation}
\Gamma^\mu_{\nu \sigma}= \frac{1}{2} \ g^{\mu \eta} \ \left(\frac{\partial
g_{\sigma \eta}}{\partial x^\nu} + \frac{\partial g_{\nu
  \eta}}{\partial x^\sigma} - \frac{\partial g_{\nu
  \sigma}}{\partial x^\eta} \right) 
\end{equation}
are the affine connection
coefficients, 
\begin{equation}
 P^\mu \equiv \frac{dx^\mu}{d\lambda} 
\label{pmu}
\end{equation}
is the four-momentum, and $g_{\mu \nu}$ is the metric tensor. Substituting (\ref{derivadacovariante}) and  (\ref{pmu}) into  (\ref{Boltzmann-coordinates}) we obtain
\begin{equation}
\left(P^\mu \frac{\partial }{\partial x^\mu} - \Gamma^\sigma_{\mu \nu}
P^\mu P^\nu  \frac{\partial}{\partial P^\mu} \right) f(x^\mu,P^\mu) = C[f] \, .
\label{secret}
\end{equation}
The pre and post collisional momentum four-vectors are connected by
energy-momentum conservation. However, it has long been known that
active neutrino
interactions on the cosmic neutrino background can be safely
neglected~\cite{Weiler:1982qy}. For simplicity, hereafter we assume that
sterile neutrino interactions can also be neglected. With this in
mind, we consider a gas of collision-free particles.

After production, the sterile neutrinos $\nu_s$ travel over
cosmological 
distances before their arrival at Earth. The flavor composition at
Earth is altered by neutrino oscillations, which are due to each
neutrino flavor state being a superposition of propagation states
$\nu_a$,
\begin{equation}
  |\nu_\alpha\rangle = \sum_{\alpha} U^*_{\alpha a} |\nu_a \rangle \,,
\end{equation}
where $U_{\alpha a}$ is an element of the mixing matrix $\mathbb U$
that connects the flavor and propagation states~\cite{GonzalezGarcia:2007ib}.  For the 3+1 scenario
under consideration $\mathbb U$ is a $4 \times 4$ unitary mixing with
16 degrees of freedom: 6 mixing angles and 3 Dirac phases~\cite{Barger:1998bn,Dutta:2001sf,Arguelles:2019tum, Ahlers:2020miq}. Unitarity
ensures conservation of the total number of neutrinos of all
flavors. Transitions from flavor $|\nu_\alpha\rangle$ to
$|\nu_\beta\rangle$ (or from $|\overline \nu_\alpha \rangle$ to $|\overline
\nu_\beta \rangle$) can only be described by their
oscillation-averaged transition probability, which is found to be $P_{\alpha \beta} =
  \sum_a |U_{\alpha a}|^2 |U_{\beta
    a}|^2$~\cite{Learned:1994wg}. All in all, the neutrino fluxes of
  different flavors at IceCube is  given by 
\begin{equation}
\left. \frac{d\Phi_{\nu_\alpha}}{dE} \right|_\oplus
=  P_{\alpha s} \frac{d \Phi_{\nu_s}}{dE} = \sum_a 
|U_{\alpha a}|^2 \ |U_{sa}|^2 \ \frac{d\Phi_{\nu_s}}{dE} \,,
\end{equation}
after all terms depending on mass squared differences are averaged out over cosmological distances. Assuming a {\it general} unitary mixing in the 3+1 flavor scenario
such that the mixing parameters relevant for high-energy neutrinos are
unbounded~\cite{Ahlers:2020miq}, Monte Carlo samples show that for a source of 100\% sterile neutrinos there is a
maximum of 75\% transformed into active flavors after oscillations, i.e., one is left with
a minimum of 25\% sterile neutrinos at Earth~\cite{Markus}. The
effective fraction of active neutrinos on Earth resulting from
oscillations of sterile neutrinos will be taken as a free parameter of
the model, but constrained to be less than 75\%.

For each neutrino mass state $m_a$, we identify \mbox{$\lambda = \tau_a/m_a$,} where $\tau_a$ is the proper time. The norm of the four-momentum reads 
\begin{equation}
{\cal P}^2 \equiv
g_{\mu \nu} P^\mu P^\nu = -E^2/c^2 + p^2 = -m_a c^2,
\label{massshell}
\end{equation}
with  $E^2/c^2 = -g_{00} (P^0)^2$ and $p^2 \equiv g_{ij}P^iP^j$, and
where $E$ is
the neutrino energy and $p$ its physical (or proper) momentum. Because
of the mass shell relation (\ref{massshell}) without loss of generality we take
only the spatial momentum components $P^i$ as independent
variables~\cite{Webb:1985}. The distribution function in the restricted phase space is
then given by
\begin{equation}
\left( P^\mu \frac{\partial}{\partial x^\mu} - \Gamma^i_{\nu \sigma} P^\nu
 P^\sigma \frac{\partial}{\partial P^i} \right) f(x^\mu,P^i) = 0 \, .
\label{seis}
\end{equation}
For a flat Friedmann-Robertson-Walker spacetime, the line element reads 
\begin{equation}
  ds^2 = - c^2 dt^2 + a^2(t) \left[ dr^2 + r^2 (d \theta^2 + sin^2 d\phi^2) \right] \,,
\end{equation}
where the time coordinate ($t$) indicates the cosmic time and the spatial
coordinates ($r, \theta, \phi$) are comoving coordinates. For each $t$, the spatial slices
are maximally symmetric, with $a(t)$ the scale factor that gauges how the
distance between two points scales with time. The Christoffel symbols
are given by $\Gamma^0_{00} = 0$, $\Gamma^0_{0i} =0$, $\Gamma_{ij}^0 =
\delta_{ij} a \dot a/c$, and $\Gamma^{i}_{0j} = \delta^i_j H/c$, where
$H = \dot{a}/a$ is the Hubble parameter.

From now on, we take $c=1$ to simplify notation. Substituting the coefficients of the affine connection into (\ref{seis}) we have
\begin{equation}
 \left(E \ \frac{\partial}{\partial t} + P^i \frac{\partial}{\partial x^i} -
   2 E  H P^i \frac{\partial}{\partial P^i} \right) f(x^\mu, P^i) = 0 \, ,
 \label{ocho}
\end{equation}
where the factor of 2 reflects that the connection coefficients are
symmetric in the lower indices.

Nevertheless, because of homogeneity of (Friedmann-Robertson-Walker)
spacetime, $f$ cannot depend on the spacial coordinates $x^i$, and so the second term
in (\ref{ocho}) is identically zero. Likewise, because of isotropy  the phase space function can only depend on the
absolute value of the momentum $P^2 =  \delta_{ij} P^iP^j$. This
implies that (\ref{ocho}) takes the form
\begin{equation}
 \left( \frac{\partial}{\partial t} - 2 H P \frac{\partial}{\partial
     P} \right) f(t,P) = 0 \, ,
\label{nueve}
\end{equation}
or using the proper momentum can be rewritten as
\begin{equation}
\left( \frac{\partial}{\partial t} -  H p \frac{\partial}{\partial
     p} \right) f(t,p) = 0 \, ,
\label{catorce}
\end{equation}  
where we have made use of the fact that for a generic function $f=f(x^2)$ it follows that
$x^i \ (\partial f/\partial x^i) = x \ (\partial f/\partial x)$, with $x^2 \equiv \delta_{ij}x^i x^j$~\cite{Piattella:2018hvi}.

Because of isotropy over the momentum space the neutrino number density relates to the phase space distribution according to
\begin{equation}
  n_\nu (t,p) \ dp = \frac{g}{(2 \pi)^3} 4\pi p^2 \ f (t,p) \  dp ,
\end{equation}
where we have allowed for $g=1/2$ internal degrees of freedom of Weyl spinors~\cite{Anchordoqui:2018qom}. Multiplying (\ref{catorce}) by $g d^3p/(2 \pi)^3$ it follows that
\begin{equation}
    \left(\frac{\partial}{\partial t} + 3 H \right) n_\nu (t,p) = 0
    \,,
    \label{dieciseis}
  \end{equation}
where the second term has been integrated by parts:
\begin{equation}
- H \frac{g}{(2\pi)^3} \int 4 \pi p^3 \ dp \ \frac{\partial f}{\partial
  p}  = 3H \frac{g}{(2 \pi)^3} \int   4 \pi p^2 dp  \ f .
  \end{equation}
Even under the assumption that neutrinos propagate unscathed, the
neutrino energy is redshifted by a factor of $(1 + z)$ so we must
correct (\ref{dieciseis}) to account for the adiabatic energy
losses ($E^{-1} dE/dt = H$) and the source term. Introducing these two terms, (\ref{dieciseis}) can be  rewritten as  
\begin{equation}
\left(\frac{\partial }{\partial t}  + 3 H \right) n_{\nu_\alpha} (t,E) =
\frac{\partial}{\partial E} [HE n_{\nu_\alpha} (E,t)] + {\cal Q}_{s}(t,E) \,
P_{\alpha s}\,,
\label{completeDE}
\end{equation}
where we have assumed that $m_{\rm dc} \gg m_a$ to adopt the
ultra-relativistic approximation  $E \approx p$, and where the source
term ${\cal Q}_{s} (t,E)$ describes the change of the net neutrino number
density due to $\chi_{\rm dc} \to \nu_s \bar \nu_s $ decay~\cite{Berezinsky:2005fa}.

\section{Bump hunting}
\label{sec:4}

We now turn to a comparison of the predictions of our scenario with  the IceCube neutrino data. Following our previous study~\cite{Anchordoqui:2015lqa}, we set $\tau_{\rm dc} \simeq 6
\times 10^{15}~{\rm s}$ and fix the fraction of $\chi_{\rm dc}$ particles to saturate the cosmological bound
(\ref{Gammaflimit}), yielding 
$F = 4 \times 10^{-6}$. With the assumed 2-body decay ($N_\nu = 2$) of $\chi_{\rm dc}$, the produced neutrino is monoenergetic,
with energy $\varepsilon = m_{\rm dc}/2$. The neutrino energy distribution
from $\chi_{\rm dc}$ decay is given by $dN_\nu/dE = N_\nu \ \delta (E - \varepsilon)$.

The source term takes the form
\begin{equation}
{\cal Q}_{s} (t,E) =  \frac{n_{\rm dc}(t)}{\tau_{\rm dc}} \,  \frac{d
  N_\nu}{dE} \,,
\end{equation}  
where
\begin{equation}
  n_{\rm dc} (t) = Y_{\rm dc} \ s(t) \ e^{-t/\tau_{\rm dc}}
\end{equation}
is the number density of $\chi_{\rm dc}$, $s(t)$ is the entropy density with $s(t_0) \simeq 2.9 \times 10^3~{\rm cm}^{-3}$, and
\begin{equation}
  Y_{\rm dc} = 3.6 \times 10^{-9} \ \frac{F \ \Omega_{\rm dm} h^2}{m_{\rm dc}/{\rm GeV}} 
\label{yield}
\end{equation}
 is the comoving number
density at the CMB epoch.

We solve (\ref{completeDE}) using the Green's function method, with
$G_{\alpha s} (t',E';t,E)$ satisfying
\begin{widetext}
\begin{equation}
  \left(\frac{\partial}{\partial t} + 2H - HE \frac{\partial}{\partial
      E} \right) G_{\alpha s} (t', E'; t, E) = \delta_{s \alpha} \ \delta
  (t'-t) \ \delta (E' - E) \quad {\rm and} \quad
  G_{s \alpha} (t',E';t,E)_{t'>t} = 0 \, .
  \end{equation}
Then,
  \begin{equation}
    G_{\alpha s} (t',E';t,E) =\Theta(t-t') \left\{{\cal K} e^{-\int_{t}^{t'} (-2H) dt''} + \int
    e^{-\int_{t}^{t'} (-2H) dt'' }  \ dt' \ \delta_{s \alpha } \, \delta
    \ (E' -
    a(t) E/a(t')) \right\}\,,
  \end{equation}
 where $a(t)$ is the scale factor at cosmic time $t$ and ${\cal K}$ the
 integration constant~\cite{Ema:2014ufa}. The integral in the exponential is given by  
 \begin{equation}
 \int_t^{t'} 2 H dt'' = \int_t^{t'} 2 \frac{da/dt''}{a} dt'' =  \ln
 \left(\frac{a(t')}{a(t)} \right)^2 \, . 
\end{equation}
Now, without loss of generality we set ${\cal K}=0$ to obtain
\begin{equation}
    G_{s \alpha}(t',E';t,E) =\Theta(t-t') \int
  \left(\frac{a(t')}{a(t)} \right)^2  dt' \, \delta_{\alpha s} \, \delta (E' -
    a(t) E/a(t')) \, .
\end{equation}
The density of neutrinos of flavor $\alpha$ at IceCube is found to be
  \begin{equation}
    n_{\nu_\alpha} (t_0, E) = \int_{t_0}^{\cal T} dt' \int dE' \ {\cal
      Q}_s(t', E') \
    P_{s\beta} \ G_{\beta \alpha} (t',E'; t,E) \, ,
  \end{equation}
where
\begin{equation}
 {\cal T}  = \frac{1}{H_0} \int_0^{[1+z({\cal T})]^{-1}} \frac{a \ da}{\sqrt{\Omega_\Lambda
          a^4 + \Omega_{m} a + \Omega_{r}}}   
     \simeq \frac{2}{3H_0\sqrt{\Omega_\Lambda}} \sinh^{-1} \Biggl[\sqrt{\frac{\Omega_\Lambda}{\Omega_m}}\Bigl(\frac{E}{\varepsilon}\Bigr)^{3/2}\Biggr]
\simeq A\Bigl(\frac{E}{\varepsilon}\Bigr)^{3/2} \ .  
          \label{calT}
\end{equation}
\end{widetext}
In the solution to the integral in (\ref{calT}), we neglect decays prior to recombination so we can omit $\Omega_r$. We use
$a_{\rm max}=[1+z({\cal T})]^{-1}=E/\varepsilon$. 
Given  
$h=0.6766$, $\Omega_\Lambda=0.6889$ and $\Omega_m=0.3111$ \cite{Aghanim:2018eyx}, the constant $A$ is
\begin{equation}
    A = \frac{2}{3 H_0\sqrt{\Omega_m}}= 5.45\times 10^{17}\ 
    {\rm s}\ .
\end{equation}
Note that $H({\cal T})\simeq H_0\sqrt{\Omega_m}[1+z({\cal T})]^{3/2}$ when matter dominates, so
$[H({\cal T})]^{-1}\simeq 8.17\times 10^{17}  (E/\varepsilon)^{3/2}$ s for $[1+z({\cal T})]\gg 1.30$
in the matter dominated era.

The density of neutrinos of flavor
$\alpha$ at Earth is 
given by
\begin{equation}
      n_{\nu_\alpha} (t_0,E )=  
      \frac{N_\nu Y_{\rm dc} s(t_0)}{\tau_{\rm dc} E}  
\frac{e^{-{\cal T}/\tau_{\rm dc}}}{H({\cal T})} \ P_{s\alpha}  \,.
\end{equation}
Finally, the all-flavor flux of active SM neutrinos at IceCube is found to be
 \begin{eqnarray}
     E^2 \Phi_{({\nu+\bar \nu})_{\rm SM}} (t_0,E) & = &\sum_{e,\mu,\tau}  
                                                   \frac{c}{4 \pi} E^2\, n_{\nu_\alpha} (t_0,E) \nonumber \\
      & = & \frac{c}{4\pi} \frac{N_\nu Y_{\rm dc} s(t_0)}{\tau_{\rm dc} }  
 \frac{E\, e^{-{\cal T}/\tau_{\rm dc}}}{H({\cal T})} \ \varkappa\,, 
\label{flux}
 \end{eqnarray}
where 
\begin{equation}\varkappa  = \sum_{\alpha= e,\mu,\tau} P_{\alpha s}= \sum_{\alpha,a }
|U_{\alpha a}|^2 \ |U_{sa}|^2\,.
\end{equation}
Equation~(\ref{flux}) can be written approximately as 
\begin{equation}
      E^2 \Phi_{(\nu+\bar{\nu})_{\rm SM}} \simeq {\cal N}(F\Gamma_{\rm dc}\varkappa) \ \left(
    \frac{E}{\varepsilon}\right)^{5/2}\exp\Biggl[-\frac{A}{\tau_{\rm dc}}\left(\frac{E}{\varepsilon}\right)^{3/2}\Biggr]
    \label{eq:flux2}
\end{equation}
where \begin{equation}
    {\cal N}= 2.43\times 10^{21} \frac{N_\nu}{2} \ {\rm \frac{GeV}{cm^2\, sr}}\,,
\end{equation}
given $\Omega_{\rm dm}h^2=0.1193$ \cite{Aghanim:2018eyx}. The maximum of the distribution of (\ref{eq:flux2}) occurs at
\begin{equation}
    E_{\rm peak} \simeq \frac{m_{\rm dc}}{2}\Biggl(\frac{5\tau_{\rm dc}}{3A}\Biggr)^{2/3}
    = \frac{m_{\rm dc}}{2}\Biggl(\frac{\tau_{\rm dc}}{3.27\times 10^{17}\ {\rm s}}\Biggr)^{2/3}\,.
    \label{eq:find-tau}
\end{equation}
The IceCube data fix the location of the peak of $E^2\Phi_
{(\nu+\bar{\nu})_{\rm SM}}$ \cite{Abbasi:2020jmh}. A range of mass and lifetime combinations yield a peak in the energy bin
with the highest $E^2\Phi_{(\nu+\bar{\nu})_{\rm SM}}$, namely, for $8.36\times 10^4 \leq E/{\rm GeV} \leq 1.86\times 10^5$, an average of $1.37\times 10^5~{\rm GeV}$. 

\begin{figure}[tbp]
\postscript{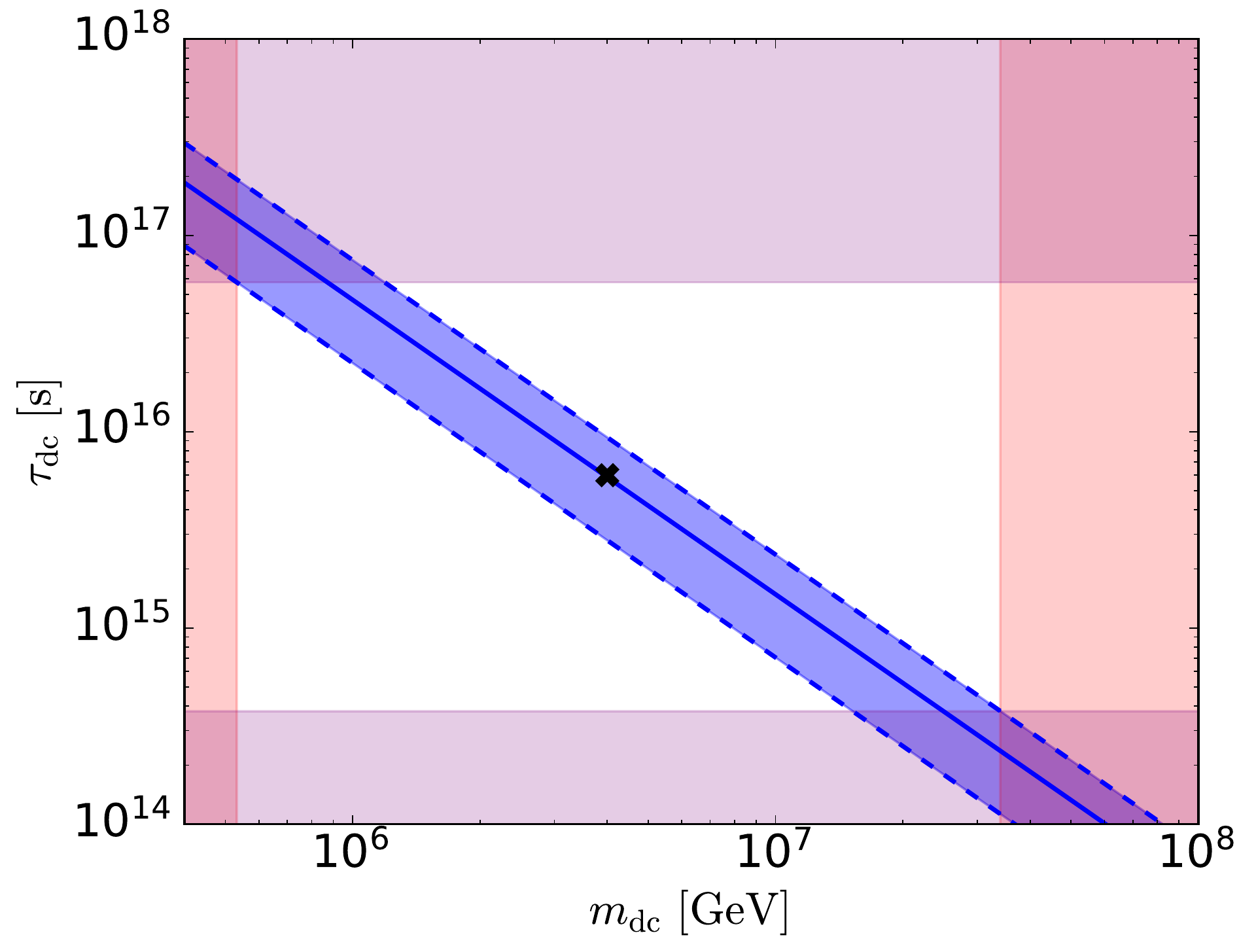}{0.9}
\caption{The lifetime $\tau_{\rm dc}$ versus $m_{\rm dc}$ for $E_{\rm peak}=1.37\times 10^5$ GeV (solid line), $8.36\times 10^4$ GeV (lower dashed line) and $1.87\times 10^5$ GeV (upper dashed line). The shaded purple regions are excluded based on the requirement that $\chi_{\rm dc}\to \nu_s\bar{\nu}_s$ decays contribute less than 1\% of the peak of $E^2\Phi_{(\nu+\bar{\nu})_{\rm SM}}$, so the shaded red regions correspond to excluded values of $m_{\rm dc}$, accounting for the range of $E_{\rm peak}$. The cross marks the fiducial point of our analysis. \label{fig:mass-lifetime}}
\end{figure}

For $E_{\rm peak}=1.37\times 10^5$~GeV and $\tau_{\rm dc}=6\times 10^{15}$~s, $m_{\rm dc}=4\times 10^6$~GeV. Note, $\varepsilon/E_{\rm peak}=(1+z_{\rm peak})=14.6$ for this value of $m_{\rm dc}$. For larger masses, the lifetime is shorter, for example, $\tau_{\rm dc}=3.8\times 10^{14}$ s for $m_{\rm dc}=2.5\times 10^7$ GeV.
For lighter masses, the lifetimes approach the age of the universe, for example, $m_{\rm dc}=10^6$ GeV corresponds to $\tau_{\rm dc}=4.7\times 10^{16}$ s.
Figure \ref{fig:mass-lifetime} shows $\tau_{\rm dc}$ as a function of $m_{\rm dc}$ for $E_{\rm peak}=1.37\times 10^5$ GeV (solid line)
and for $E_{\rm peak}= 8.36\times 10^4-1.87\times 10^5$ GeV (shaded blue band). The cross marks the location of the fiducial values $\tau_{\rm dc}=6\times 10^{15}$~s and $m_{\rm dc}=4\times 10^6$~GeV

\begin{figure}[tbp]
\postscript{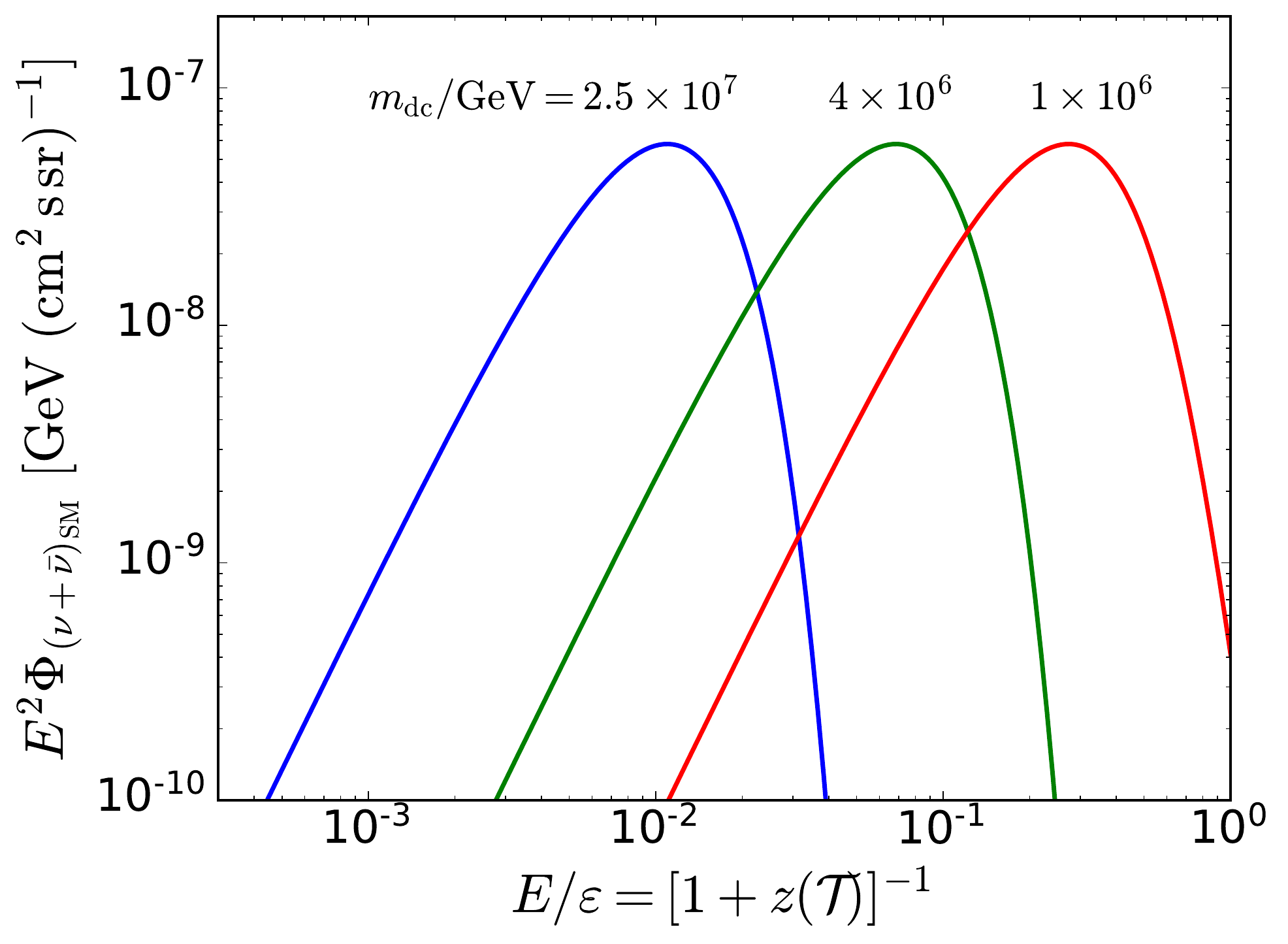}{0.9}
\caption{The quantity $E^2\Phi_{(\nu+\bar{\nu})_{\rm SM}}$ as a function of $E/\varepsilon$ assuming $E_{\rm peak}=1.37\times 10^5$ GeV and the decay of $\chi_{\rm dc}$ to sterile neutrinos contributes $\max (E^2 \Phi_{({\nu+\bar \nu})_{\rm SM}})=5.80\times 10^{-8}$ GeV/(cm$^2$s\,sr).\label{fig:flux-y}}
\end{figure}

The flux of SM neutrinos from $\chi_{\rm dc}\to \nu_s\bar{\nu}_s$, scaled by $E^2$, as a function of $E/\varepsilon$ is shown in Fig.~\ref{fig:flux-y}. Because of the form of  (\ref{eq:flux2}), the shape is the same for any $\varepsilon=m_{\rm dc}/2$ and lifetime, shifted in $E/\varepsilon$ because of $m_{\rm dc}$. In Fig.~\ref{fig:flux-y}, the flux has been normalized so that the peak of the $\chi_{\rm dc}\to \nu_s\bar{\nu}_s$ contribution is $E^2 \Phi_{({\nu+\bar \nu})_{\rm SM}} = 5.80\times 10^{-8}$ GeV/(cm$^2$s\,sr). The lifetime is chosen from (\ref{eq:find-tau}) with $E_{\rm peak}=1.37\times 10^5$~GeV.

As Figs.~\ref{fig:mass-lifetime} and \ref{fig:flux-y} show, a larger $m_{\rm dc}$ corresponds to a shorter lifetime and to the correspondingly larger redshifts to the neutrino energy $\varepsilon$ at production.
If the lifetime is too short, $\chi_{\rm dc}$ decays will occur before recombination. The requirement that the $E^2\Phi_{({\nu+\bar \nu})_{\rm SM}}$ be $<1\%$ of the peak value from $\chi_{\rm dc}$ decays at $z=1100$ corresponds to $\tau_{\rm dc}>3.8\times 10^{14}$ s. 

For smaller masses, we find an upper bound on the lifetime of $\tau_{\rm dc}^{\rm max}=5.8\times 10^{16}$ s if we require that for $z=0$, $E^2\Phi_{({\nu+\bar \nu})_{\rm SM}}$ be $<1\%$ of the peak value from $\chi_{\rm dc}$ decays. The two exclusion regions are shown by shaded purple bands in Fig. \ref{fig:mass-lifetime}. The shaded red bands correspond to excluded $m_{\rm dc}$ values, based on (\ref{eq:find-tau}), accounting for the bin width of $E_{\rm peak}$ that corresponds to the blue band. In what follows, we assume $E_{\rm peak}=1.37\times 10^5$~GeV. 

For the $\chi_{\rm dc} $ decay width in terms of an effective coupling $g_{\rm eff}$, $\Gamma_{\rm dc}=g_{\rm eff}^2 m_{\rm dc}/(16\pi)$, the lifetimes and masses considered here require $g_{\rm eff}\sim 5\times 10^{-23}$.
Given $\tau_{\rm dc}^{\rm max}= 5.8\times  10^{16}$ s, the bound on $F\Gamma_{\rm dc}$ in (\ref{Gammaflimit})
translates to 
\begin{equation}
\label{eq:flim2}
    F<4.14\times 10^{-5}\, ,
\end{equation}
several orders of magnitude lower than from the more general constraint on $F$ from CMB measurements.

With the approximate formulas for ${\cal T}$ and $H({\cal T})$ and the relation between $m_{\rm dc}$ and $\tau_{\rm dc}$, one can show that the quantity $F\varkappa$ scales with $m_{\rm dc}$ as
\begin{equation}
    F\varkappa\simeq 6.0 
    \times 10^{-10}\, \Biggl(\frac{m_{\rm dc}}{4\times 10^6\ {\rm GeV}}\Biggr)\,.
    \label{eq:fkappa}
\end{equation}
For the mass range discussed here, $8.7\times 10^5 \leq  m_{\rm dc}/{\rm GeV} \leq   2.5\times 10^7$,  we find $1.3\times 10^{-10} \leq F\varkappa \leq 3.8\times 10^{-9}$. Given (\ref{eq:flim2}),
$\varkappa>3.2\times 10^{-6}$.
For our fiducial values, $m_{\rm dc}=4\times 10^6$ GeV,
$\tau_{\rm dc}=6\times 10^{15}$~s, and $F=4\times 10^{-6}$ we obtain
$Y_{\rm dc} \simeq 4.3 \times 10^{-22}$ and $\varkappa = 1.5\times
10^{-4}$. This value of $\kappa$ can be achieved in a
  scenario with $|U_{s4}|^2 \simeq 1$ and $|U_{\alpha 4}|^2\lesssim 1.5\times
  10^{-4}$ for $\alpha=e,\mu,\tau$, values well below
  experimental constraints on active-sterile mixing~\cite{Dentler:2018sju} and much smaller than $|U_{e3}|^2\simeq
  2\times 10^{-2}$. For $\varkappa\sim 10^{-4}$ and 
$m_{\nu_s}\sim 1$ eV, active-sterile oscillations in cosmology will
not violate bounds on the effective number of neutrino species, as
discussed in, e.g., Ref.~\cite{Gariazzo:2019gyi}. 

\begin{figure}[tbp]
\postscript{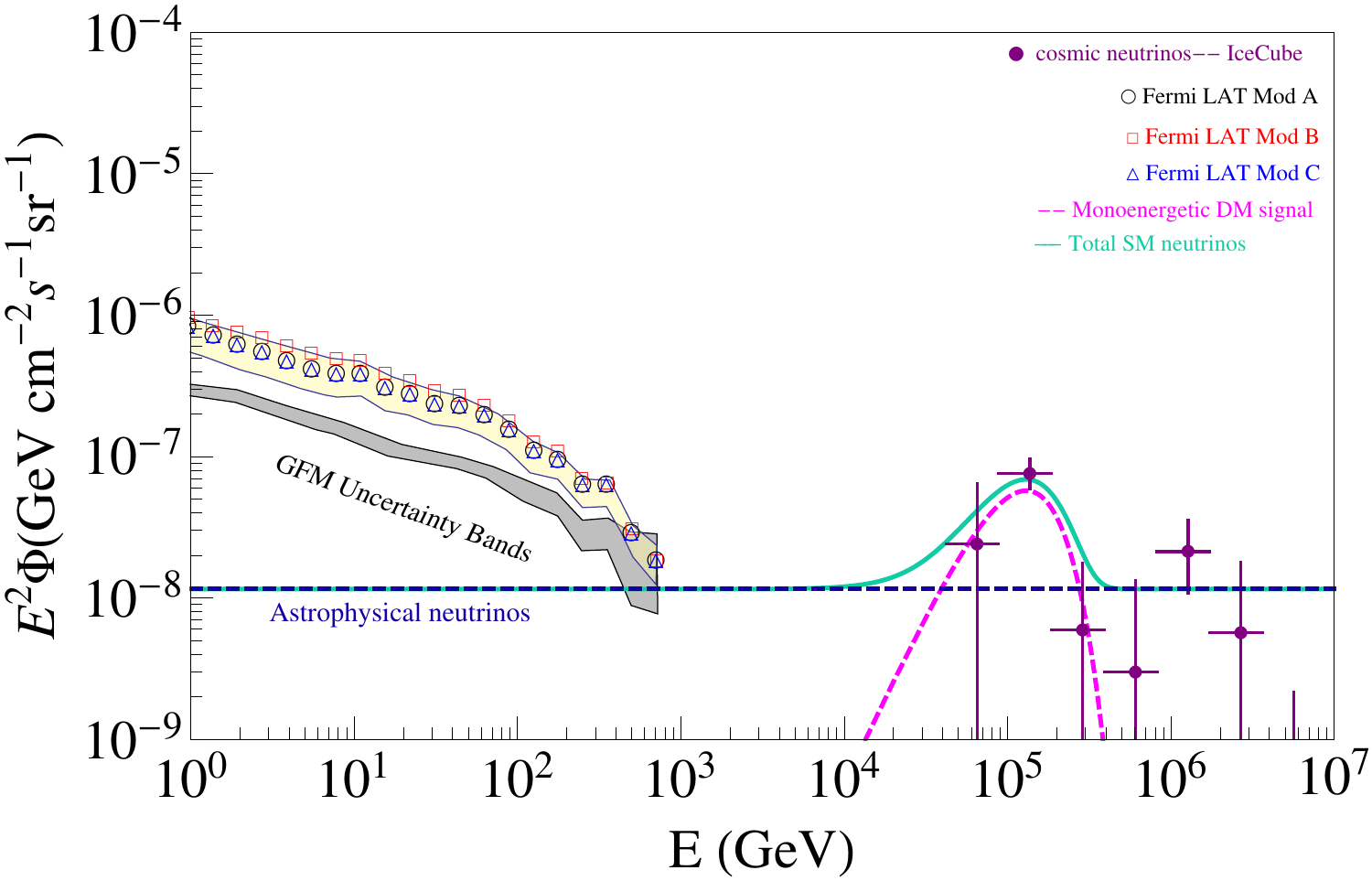}{0.9}
\caption{
The total extragalactic $\gamma$-ray background (open symbols) reported by the Fermi  Collaboration \cite{Ackermann:2014usa} and the diffuse astrophysical flux (solid symbols) reported by the IceCube Collaboration (frequentist statistcal analysis 7.5~yr data)~\cite{Abbasi:2020jmh}. The $\gamma$-ray result is shown for three foreground models, with the yellow band showing background modeling uncertainties and the gray band showing the cumulative intensity from resolved Fermi sources at latitudes $|b| > 20^\circ$. The SM diffuse astrophysical neutrino flux is shown with the horizontal dashed line, 
and the SM neutrino flux from dark matter decays is shown with the magenta dashed line. 
\label{figura}}
\end{figure}

In Fig.~\ref{figura} we show the $\gamma$-ray flux and the monoenergetic all-flavor 
flux of active SM neutrinos, smeared by redshift evolution, together with the diffuse astrophysical neutrino flux superimposed on IceCube data. 
The horizontal line in Fig.~\ref{figura} indicates the flux of
astrophysical neutrinos produced via pion decay, which 
translates to equal fluxes of neutrino flavors and is taken to 
saturate the Fermi flux, given by
\begin{equation}
\frac{\Phi_{(\nu +\bar \nu)_{\rm SM,astro}}(E)}{{\rm GeV} \ {\rm cm}^{2} \
{\rm sr} \ {\rm s}}
 = 1.5 \times 10^{-18}
 \left(\frac{E}{10^5~{\rm GeV}}\right)^{-2} \, .
 \label{treintaytres}
 \end{equation}
A cutoff (or steepening) of the astrophysical neutrino spectrum around
$E \sim 10^{7}~{\rm GeV}$ is
necessary to accommodate the non-observation of $\nu_e$ or
$\nu_\tau$ events above this energy~\cite{Aartsen:2020aqd}. The astrophysical neutrino flux adopted herein is consistent at the $1\sigma$ level with the muon neutrino flux reported by the IceCube Collaboration~\cite{Aartsen:2015rwa}. The (magenta) dashed line is
obtained for our fiducial parameters, or alternatively, with a combination of parameters that satisfy (\ref{eq:find-tau}) and (\ref{eq:fkappa}).
The active flavor fractions of this all-flavor neutrino flux at Earth are essentially unconstrained, as
discussed in~\cite{Ahlers:2020miq}.

\section{Related Phenomenology}

\label{sec:5}

Sterile neutrinos relevant to the MiniBooNE~\cite{MiniB} and LSND~\cite{LSND} anomalies have masses $1 \alt m_s/{\rm eV} \alt 10$ and active-sterile neutrino mixing with an amplitude at the level of 0.01. IceCube places constraints on this $3+1$ oscillation framework by searching for Earth matter effects in oscillations of atmospheric neutrinos with energies between 200~GeV and 10~TeV~\cite{Aartsen:2020fwb}. However, for the 100~TeV energies of our scenario, no such matter effects are observable at IceCube.
This conclusion also applies for sterile neutrino masses above 10~eV~\cite{Blennow:2018hto}.

Proposed neutrino detection experiments along the LHC beamline such as \textsc{Faser} \cite{Feng:2017uoz,Ariga:2018pin,Ariga:2019ufm} and \textsc{ Faser}$\nu$ \cite{Abreu:2019yak,Abreu:2020ddv}, \textsc{ Xsen} \cite{Buontempo:2018gta,Beni:2019pyp} and \textsc{Snd\@LHC}
\cite{Ahdida:2020evc} have the potential to probe active-sterile mixing. Neutrinos that travel down the beamline are produced by forward pions and kaons that decay outside the detector, from $W$ and $Z$ decays, and from prompt decays of charmed hadrons. Indeed, in the far forward region, the dominant source tau neutrinos is from $D_s^\pm$ decays \cite{Beni:2019gxv}, making $\nu_\tau\to\nu_s$ oscillations the cleanest signal. Even so, large uncertainties in forward charm production kinematic distributions and cross sections mean that until charm production calculations are better refined with new forward production data, signals of $\nu_\tau\to\nu_s$ oscillations would be via observations of spectral distortions \cite{Bai:2020ukz}. 
With baselines of these far-forward experiments on the scale of 500 m, a characteristic sterile neutrino mass scale of 10's of eV will show oscillation dips for neutrino energies of order $\sim 150$ GeV if mixing angles are large enough.
To be observable, sufficiently large values of $\varkappa$ are required, in particular, that $\mid U_{\tau 4}\mid^2\simeq 0.15$, close to the maximum allowed value \cite{Dentler:2018sju}. To avoid the cosmological constraints on such a large mixing angle, additional ``secret interactions'' ({\it viz.} $C[f] \neq 0$) could be invoked, as in, e.g., Refs.~\cite{Chu:2015ipa,Archidiacono:2016kkh,Jeong:2018yts}.

\section{Conclusions}
\label{sec:6}

We propose a novel explanation of the spectrum of cosmic neutrinos on the basis of cold dark matter which decays into sterile neutrinos that, after oscillations, produce the bumpy signal observed in IceCube data. The dark matter particle decays via $\chi_{\rm dc} \to \nu_s \overline \nu_s$ with a branching fraction of essentially unity, and thereby eliminates the IceCube-Fermi tension. The scenario features interesting phenomenology which can be summarized as follows:
\begin{itemize}[noitemsep,topsep=0pt]
\item The total decay width of the dark matter particle satisfies $H_0 < \Gamma_{\rm dc} \alt H(z_{\rm LS})$. The lower limit ensures that most of the $\chi_{\rm dc}$ particles have disappeared by $z = 3$, so monoenergetic neutrinos are not found in IceCube searches of dark matter decay in the Galactic center~\cite{Aartsen:2018mxl}. The upper limit
on  $\Gamma_{\rm dc}$ ensures that the portion of the overall dark matter abundance that has been depleted by decays prior to last scattering is negligible, avoiding modifications of $\Lambda$CDM predictions from big bang nucleosynthesis or CMB phenomenology.
\item A subdominant CDM component decaying into dark radiation after recombination depletes dark matter density at low redshifts reducing the power of the CMB lensing effect, which is at odds with  Planck data. This sets an upper limit of
$F \,  \Gamma_{\rm dc}  < 2.25 \times
  10^{-5}~{\rm Gyr}^{-1}$ at 95\% CL~\cite{Nygaard:2020sow}. Normalization to the IceCube flux yields a constraint on $\kappa F \,  \Gamma_{\rm dc}$ after sterile-active neutrino oscillations, with mixing amplitude $\propto \kappa$.
\item The thermalization of extra light species in the early universe modifies the energy density of radiation. The
presence of additional relativistic degrees of freedom can be characterized by the number of equivalent light neutrino species 
$N_{\rm eff} \equiv (\rho_{\rm R} - \rho_\gamma)/\rho_{\nu}$  in units of the density of a single Weyl neutrino $\rho_\nu$, where
$\rho_{\rm R}$ is the total energy density in relativistic particles
and $\rho_\gamma$ is the energy density of
photons~\cite{Steigman:1977kc}. For $1 \alt m_s/{\rm eV} \alt 10$ and our fiducial value $\varkappa = 1.5 \times 10^{-4}$, the mixing is not large enough to allow a full energy transfer between the sterile and the active states~\cite{Gariazzo:2019gyi}. Indeed, due to incomplete thermalization the  model prediction for $N_{\rm eff}$ is virtually indistinguishable from that of three families of massless SM neutrinos
($N_{\rm eff} = 3.046$~\cite{Mangano:2005cc}) in  $\Lambda$CDM cosmology.
\item The allowed range for the sterile neutrino mass encompasses the mass scale  of the long-standing anomalies in  MiniBooNE~\cite{MiniB} and LSND~\cite{LSND}. However, for the preferred region of active-sterile mixing parameters from short-baseline neutrino experiments, the sterile neutrino is fully thermalized ($N_{\rm eff} \simeq 4$) and therefore in strong tension with the upper limit derived by the Planck Collaboration: $N_{\rm eff} = 3.12^{+0.25}_{-0.26}$ at the 95\%~CL~\cite{Aghanim:2018eyx}. 
\item The energy of the sterile neutrino flux is outside the range of oscillation searches for sterile neutrinos at IceCube~\cite{Aartsen:2020fwb}.
\item Proposed neutrino detection experiments  along  the  LHC  beamline will be sensitive to active-sterile neutrino mixing if $\kappa$ is sufficiently large~\cite{Bai:2020ukz}. To avoid the cosmological constraints on such a large mixing with the hidden sector, a collision term ($C[f] \neq 0$) should be added to (\ref{secret}) to account for secret neutrino interactions~\cite{Chu:2015ipa,Archidiacono:2016kkh,Jeong:2018yts}.
\end{itemize}
In summary, we have proposed a new scenario to explain IceCube data. This scenario can be confronted with future data from IceCube-Gen2~\cite{Aartsen:2020fgd} and KM3NeT~\cite{Aiello:2020tki}.
Observation by both of these neutrino-detection facilities will be able to test whether the bumpy signal has a cosmological origin.

 \acknowledgments{L.A.A. is supported by the U.S. National Science Foundation (NSF) Grant PHY-1620661 and the National Aeronautics and Space
    Administration (NASA) Grant 80NSSC18K0464.  V.B. is supported  by the U.S. Department of Energy (DoE) Grant DE-SC-0017647.  D.M. is supported by the DoE Grant DE-SC-0010504. M.H.R. is supported by the DoE Grant DE-SC-0010113.
    T.J.W. is supported  by the DoE Grant DE-SC-0011981. T.J.W. thanks Vanderbilt University and its Physics Office for naming him Emeritus Professor of Physics, effective upon his retirement December 31, 2020. Any opinions,
    findings, and conclusions or recommendations expressed in this
    material are those of the authors and do not necessarily reflect
    the views of the NSF, NASA, or DoE.}
    
 \section*{Note added}

    After our paper was published, the authors of Ref.~\cite{Nygaard:2020sow} corrected the 95\% CL upper bounds in
    Eqs.~(\ref{eq:flim}) and~(\ref{Gammaflimit}) to
    \begin{equation}
      F < 2.62 \times 10^{-2}
    \end{equation}
    and
    \begin{equation}
     F \ \Gamma_{\rm dc}  < 5.84 \times
      10^{-2}~{\rm Gyr}^{-1}\,.
    \end{equation}
    Since our fiducial values are comfortably consistent with the revised constraints, our conclusions remain unchanged.

\end{document}